\documentclass[]{aastex62}
\submitjournal{ApJ}
\usepackage{graphicx}
\usepackage{lineno}

\usepackage{bm}
\usepackage{amsmath,amssymb,amsfonts,textcomp,array}
\usepackage{comment}
\usepackage{xcolor}
\def\lsim{\mathrel{\mathstrut\smash{\ooalign{\raise2.5pt\hbox{$<$}\cr\lower2.5pt\hbox{$\sim$}}}}}
\def\gsim{\mathrel{\mathstrut\smash{\ooalign{\raise2.5pt\hbox{$>$}\cr\lower2.5pt\hbox{$\sim$}}}}}

\def\beq{\begin{equation}\begin{aligned}}
\def\eeq{\end{aligned}\end{equation}}

\usepackage{tikz}
\usetikzlibrary{quotes,arrows.meta}
\usetikzlibrary{%
    decorations.pathreplacing,%
    decorations.pathmorphing%
}
\usetikzlibrary{decorations.markings}
\usetikzlibrary{positioning,arrows} 
\tikzset{photon/.style={decorate, decoration={snake}, draw=black},
fermion/.style={thick,draw=black, postaction={decorate},
    decoration={markings,mark=at position .5 with {\arrow[black]{triangle 45}}}},
gluon/.style={decorate, draw=black,
    decoration={coil,aspect=0}},
scalar/.style={thick,dashed,draw=black, postaction={decorate}},
higgs/.style={thick,loosely dashed,draw=black, postaction={decorate}}} 

\newcommand{\LCDM}{$\Lambda$CDM }

\begin{document}

\title{EARLY MASS VARYING NEUTRINO DARK ENERGY: NUGGET FORMATION AND HUBBLE ANOMALY}

\author{Antareep Gogoi}
\email{ag15ms096@iiserkol.ac.in}
\affiliation{Department of Physical Sciences, IISER Kolkata, WB, 741246, India}

\author{Ravi Kumar Sharma}
\email{ravi.sharma@iiap.res.in}
\affiliation{Indian Institute of Astrophysics, Bangalore, 560034, India }

\author{Prolay Chanda}
\email{pchand31@uic.edu}
\affiliation{ Department of Physics, University of Illinois at Chicago, Chicago 60607, USA}

\author{Subinoy Das}
\email{subinoy@iiap.res.in}
\affiliation{Indian Institute of Astrophysics, Bangalore, 560034, India }

\begin{abstract} 
We present a novel scenario, in which light ($\sim$ few \rm{eV}) dark fermions (sterile neutrinos) interact with a scalar field like in mass varying neutrino dark energy theories. As the $\rm{eV}$ sterile states naturally become non-relativistic before the Matter Radiation Equality (MRE), we show that the neutrino-scalar fluid develops strong perturbative instability followed by the formation of  neutrino-nuggets and the early dark energy behaviour disappears around MRE. The stability of the  nugget is achieved when the Fermi pressure balances the attractive scalar force and we numerically find the mass and radius of heavy cold nuggets by solving for the static configuration for the scalar field.  We find that  for the case when DM nugget density is sub-dominant and most  of the early DE energy goes into scalar field dynamics, it can in principle relax the Hubble anomaly. Especially when a kinetic energy dominated phase appears after the phase transition, the DE density dilutes faster than radiation and satisfy the requirements for solving $H_0$ anomaly. In our scenario, unlike in originally proposed  early dark energy theory, the dark energy density is controlled by ($\rm{eV}$) neutrino mass and it does not require a fine tuned EDE scale. We perform a MCMC analysis and confront our model with Planck + SH0ES and BAO data and find an evidence for non-zero neutrino-scalar EDE density during MRE. Our analysis shows that this model is in agreement of nearly 1.3$\sigma$ with SH0ES measurement which is $H_0 = 74.03 \pm 1.42$ km/s/Mpc.
   \end{abstract}

\section{Introduction}
\label{sec:intro}

The existence of dark matter, dark energy, and the discovery of neutrino mass have made the frontier of cosmology and particle physics both fascinating and challenging to explain. 
A sterile neutrino with mass $m \gtrsim 5\, {\rm keV} $ is a viable  warm  dark matter candidate \citep{Dodelson:1993je,Valdarnini:1998zy,Boyarsky:2008mt,Bezrukov:2009th,Petraki:2007gq,Laine:2008pg,Abazajian:2005xn,Irsic:2017ixq,Gilman:2019nap,Nadler:2020prv}. However, a lighter fermion with $m \sim {\rm eV}$ usually falls into the same dark-matter misfortune as it free-streams until a relatively late times in cosmic evolution and erases structure on small scales.  Only a tiny fraction of the dark matter abundance can be in the form of neutrinos or other light fermions which puts a stringent upper bound on neutrino mass \citep{Hannestad:2010yi,Bell:2005dr,Acero:2008rh}. On the other hand, recent data from the MiniBooNE experiment \cite{AguilarArevalo:2010wv,Aguilar-Arevalo:2013pmq,Aguilar-Arevalo:2018gpe} might indicate the existence of light sterile neutrino states (\rm{eV - 10 eV}) and also the possibility of these light states having nontrivial interactions \citep{Dasgupta:2013zpn}. There have been attempts to revive the neutrino or lighter sterile neutrino as viable dark matter candidate with nontrivial cosmological histories or exotic interactions \citep{Das:2006ht,Bjaelde:2010vt, Abazajian:2019ejt}.

 In this paper, we propose a  scenario in which at some point in the radiation dominated era (RDE),  a population of eV-mass decoupled fermions undergo a phase transition and clump into small dark matter nuggets due to the  presence of a scalar mediated fifth force. Before the transition, the neutrino-scalar fluid  behaves like early dark energy (EDE) as the field adiabatically  stays at the minimum of the effective potential, $ V_{\rm eff}$.
 Once the transition happens, fermions inside the nugget no longer free-stream and the nuggets as a whole behave like cold dark matter. Due to its interaction with a scalar $\phi$, the fermion mass is ``chameleon'' \citep{Khoury:2003aq} in nature and depends on scalar vacuum  expectation value. The dynamics of $\phi$ are controlled by $ V_{\rm eff}$ instead of just $V(\phi)$ \citep{Fardon:2003eh,Das:2005yj}, and the scalar field adiabatically tracks the minima of $ V_{\rm eff}$. 
 
 As the background fermion density dilutes due to the expansion of the universe, the minimum of $V_{\rm eff}$ is time-dependent and so is the fermion mass. We consider a scenario where the fermion mass is inversely proportional to  the scalar \textit{vev} (through the see-saw mechanism \citep{Yanagida:1979as,Yanagida:1980ve,GellMann:1980vs,Mohapatra:1979ia}) such that at a red-shift $z_F$ (before matter-radiation equality), it becomes non-relativistic  and the attractive scalar force starts to dominate over the free-streaming. In situations like this, it has been shown in \cite{Afshordi:2005ym} that the effective sound speed of perturbation of the combined fluid (scalar and non-relativistic fermions) becomes imaginary following a hydro-dynamical instability which results in nugget formation. The majority of fermions within each scalar Compton volume collapse into a nugget until the Fermi pressure intervenes and balances the attractive force.  From our numerical results we find that the scalar field inside the nugget is displaced in such a way  that the fermion mass inside the nugget is much smaller than outside, ensuring the stability of the nuggets. Along with the nugget profile's numerical solution, we also solve for the entire cosmological evolution for our scenario and obtain CMB angular power spectra. We find that the effects on CMB is more prominent when phase transition happens closer to MRE this in turn provides a conservative lower bound on transition redshift $z_F \geq 10^5 $. 
 
  As an important aspect of the nugget formation mechanism, we explore its implication for recent  Hubble tension as the neutrino-scalar fluid naturally  behaves as dark energy fluid prior to nugget formation. To resolve the recent Hubble anomaly  where local distance ladder measurements of $H_0$ disagree with the Planck measured value at $ \sim 5\sigma$ \citep{Humphreys:2013eja,Verde:2019ivm,Wong:2019kwg}, one obvious choice is to increase dark radiation before CMB.  But it is a well-known fact that  just increasing dark radiation content ($\Delta N_{\textrm{eff}}$) only partially resolves Hubble tension as it makes the high-$\ell$ CMB prediction deviate from Planck observation \citep{Riess:2016jrr,Ichikawa:2007jv,Blinov:2020hmc}.  But instead of increasing dark radiation, the appearance of an early dark energy component before CMB has been one of the most well studied and successful avenues \citep{Karwal:2016vyq,Poulin:2018dzj,Poulin:2018cxd,Agrawal:2019lmo,Braglia:2020bym,Chudaykin:2020acu,Ivanov:2020ril,Weiner:2020sxn,Hill:2018lfx,Smith:2020rxx} for solving Hubble tension though there are challenges \cite{Jedamzik:2020zmd} especially when the model is subject to recent LSS results. From the theoretical perspective, the physics of matter-radiation equality and early dark energy are completely disconnected, so that some degree of fine-tuning is needed them to appear nearly simultaneously. However in our case, the instability of the neutrino-scalar fluid naturally  happens around the MRE, which occurs due to the transition of massive neutrinos from relativistic to non-relativistic. There has been recent work \citep{Sakstein:2019fmf,Niedermann:2019olb,Niedermann:2020dwg,Niedermann:2020qbw}
where EDE has been connected to neutrino physics but in a different context than ours.
  
In this work, we qualitatively demonstrate  that the scalar field adiabatically traces the minimum of the effective potential of neutrino-scalar interaction prior to nugget formation, thus behaving like an early DE for a short duration. This early dark energy (EDE) behaviour of the neutrino-scalar fluid  increases the Hubble expansion rate  in a natural way as neutrino mass controls the period of EDE. As soon as the nugget forms, neutrino decouples from the scalar field and the scalar  no longer behaves like DE. We find that though the presence of EDE is natural in this scenario of nugget formation, if the entire energy goes from dark energy to dark matter (as in simple model of DM formation) it only partially solves the Hubble anomaly. This is expected as it is now a well established fact in the literature that the early dark energy phase should dilute at least like radiation or faster to resolve Hubble anomaly\cite{Lin:2019qug}. But as we discuss qualitattively, it is highly possible that a major fraction of the early DE phase goes into the scalar field and provides a solution to Hubble anomly, especially through a scalar field kinetic energy dominated phase following the nugget formation, similar to \cite{Alexander:2019rsc}.
Our ongoing MCMC analysis ( work in progress ) with different potential with early  mass varying neutrino dark energy model will clarify how effectively one can resolve Hubble tension in this scenario. 

The plan of the paper is as follows: In Section \ref{sec:instability}, we derive the condition for instability for a quadratic potential. In Section \ref{sec:static}, we solve for the static configuration of the scalar field and show nuggets indeed form. In Section \ref{sec:cosmology}, we derive the perturbation equation and  numerically find for CMB angular power spectra followed by discussion of Hubble anomaly in Section \ref{sec:hubble}. 
In Section \ref{subsec:constrains}, we study stability of DM nugget and applicable constraints on  $\Delta N_{\rm eff}$ bounds given by Planck and BBN results. Finally we conclude in Section \ref{sec:conclusion} and discuss future directions.

\section{How it works: Derivation of instability from effective sound speed} \label{sec:instability}

In the  original mass varying neutrino model \citep{Fardon:2003eh} the Majorana mass term of a heavy chiral fermion in dark sector was taken to be a linear function of a scalar field  $\phi$. Here we consider a more general interaction involving a function $f(\phi)$ with the following Lagrangian
\begin{equation}
{\cal L} \supset m_{D} \psi_1 \psi_2 + f(\Phi) \psi_2 \psi_2 + V(\Phi) + \mbox{H.c.},
\end{equation} 
where $\psi_{1,2}$ are fermion fields, with $\psi_2$ corresponding to the heavier mass eigenstate. Both fermion fields are written as two component left chiral spinors, $m_D$ is the Dirac mass term and $V(\phi)$ is the scalar potential. Note that if  $\psi_2$ is considerably heavier, we can integrate it out from the low energy effective theory obtaining  $m(\phi)\equiv m_{\psi_1}= m_D^2/f(\phi)$, otherwise one can obtain the mass eigenvalues by diagonalizing  the fermion mass matrix.\\ 
We adopt a simple quadratic potential $V(\phi)= m_{\phi}^{2}\phi^{2}$ and assume a Yukawa-type interaction with $f(\phi)= \tilde{\lambda} \phi$, such that the model looks like
 \begin{equation}
{\cal L} \supset m_D \psi_1 \psi_2 + \tilde{\lambda} \phi \psi_2 \psi_2 + V(\phi).
\end{equation} 
However, prior to phase transition, the dynamics of $\phi$ is controlled by an effective potential given by \cite{Fardon:2003eh}:
  \begin{equation} \label{eq:veff}
  V^{\rm eff}= \rho_{\psi_1} + m_{\phi}^{2}\phi^{2}. 
   \end{equation}
  It is instructive to note that radiative correction to any dark energy potential  
  is a common issue and will also be there for our quadratic potential. But there has been some effort to control it through a super symmetric theory of neutrino dark energy \cite{Fardon:2005wc}. 
As the first term decreases due to Hubble expansion, the minima of the potential moves to a lower value, thus increasing the mass of the lightest fermion eigenstate. 
  In our model, $\psi_1$ is taken to be a sterile neutrino of mass $\sim$ eV. As the field evolves adiabatically, the field responds to the average neutrino density, relaxing at the minimum of the effective potential. In this minimum, we see that both mass of the neutrino, $m_{\psi_1}$ and the scalar field potential, $V(\phi)$ become functions of the neutrino density,
\begin{equation}
\frac{\partial V_{\textrm{eff}}}{\partial \phi} = \left( n_{\psi_1} + \frac{\partial V}{\partial m_{\psi_1}}\right) \frac{\partial m_{\psi_1}}{\partial \phi} = 0 \Rightarrow n_{\psi_1} = -\frac{\partial V}{\partial m_{\psi_1}}.
\label{eq:four}
\end{equation}
We have replaced $\rho_{\psi_1}$ with $n_{\psi_1} m_{\psi_1}$, where $n_{\psi_1}$ denotes the number density. The right side of the eq.~(\ref{eq:four}) holds provided that $\partial m_{\psi_1}/\partial \phi$ doesn't vanish. In the non-relativistic limit, there would be no pressure contribution from the neutrinos and similar to \cite{Fardon:2003eh}, we neglect the contribution of any kinetic terms of the scalar field to the energy density, which is a good approximation as long as the scalar field is uniform and tracks the minimum of its effective potential adiabatically. Therefore, we can describe the $\phi$-$\psi_1$ fluid by the equation of state parameter:
\begin{equation}
  w \equiv \frac{\rm Pressure}{\text{Energy Density}} = \frac{-V}{n_{\psi_1} m_{\psi_1}+V} = -1 + \frac{n_{\psi_1} m_{\psi_1}}{V_{\textrm{eff}}}, \label{eq:weos}
  \end{equation}
where we used eq.~(\ref{eq:veff}) in the last step. After a few steps of calculation using eqs.~(\ref{eq:four}) and ~(\ref{eq:weos}) and the the relation between $V(\phi)$ and $m_{\psi_1}$ through $\phi$, we get,
    \begin{eqnarray}
n_{\psi_1} m_{\psi_1} &=& -\frac{\partial V}{\partial m_{\psi_1}}m_{\psi_1} = 2m_\phi^2\left(\frac{m_D^2}{\tilde{\lambda} m_{\psi_1}}\right)^2 = 2V\\
\Rightarrow w &=& -1 + \frac{2V}{2V+V} = -\frac{1}{3}.  \label{eq:wnonrel}
\end{eqnarray}
  
For perfect fluids, the speed of sound purely arises from adiabatic perturbations in the pressure and the energy density. Hence, the adiabatic sound speed, $c_a^2$ of a fluid can be purely determined by the equation of state. The adiabatic sound speed of our fluid is given by
\begin{equation}
c_{a}^{2} \equiv \frac{\dot{p}}{\dot{\rho}} = w - \frac{1}{3} \frac{\dot{w}}{(1+w)H} = -\frac{1}{3}. \label{eq:cs2}
\end{equation}
However, in imperfect fluids, dissipative processes generate entropic perturbations in the fluid and therefore we have the more general relation
\begin{equation}
    c_s^2 = \frac{\delta p}{\delta\rho}.
\end{equation}
This can also be written in terms of the contribution of the adiabatic component and an additional entropic perturbation $\Gamma$ and the density fluctuation in that instantaneous time frame \citep{Kodama:1985bj}:
\begin{equation}
    w\Gamma = (c_s^2-c_a^2)\delta.
\end{equation}

As stated before, the scalar field follows the instantaneous minimum of its potential and this minimum is modulated by the cosmic expansion through the changes in the local neutrino density. So, the mass of the scalar field, $m_\phi$ can be much larger than the Hubble expansion rate. Consequently, the coherence length of the scalar field, $m_\phi^{-1}$, is much smaller than the present Hubble length. As a consequence of this \cite{Afshordi:2005ym}, combined with eq.~(\ref{eq:cs2}), the sound speed of the neutrino-scalar fluid becomes imaginary once the neutrino becomes non-relativistic followed by instability and  formation of nuggets.
We have solved the linear perturbation equation (For details see Section \ref{sec:cosmology}) to verify this for our choice of quadratic potential. As shown in Fig.1, we compare the evolution of density perturbation  of a typical mode (smaller than the range of scalar force) in the radiation dominated era.

\begin{figure}[t]
    \centering
    
    \includegraphics[scale=0.5]{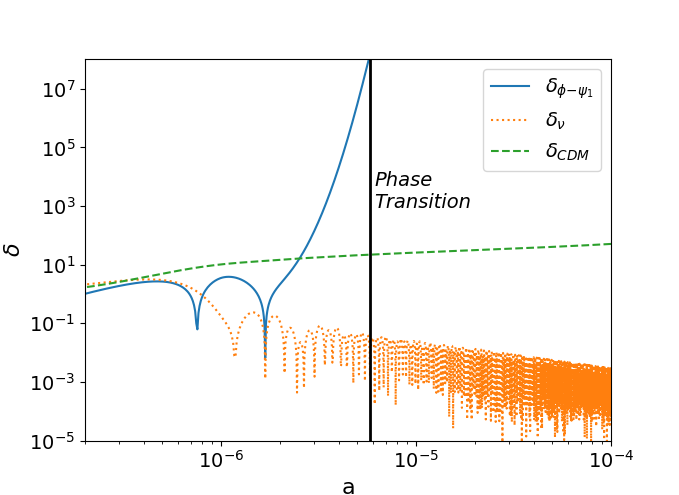}
    \caption{The evolution of density perturbation  of a typical mode (smaller than the range of scalar force) in radiation dominated era. We see that as neutrino turns non-relativistic and sound speed becomes imaginary, the perturbation in neutrino-scalar fluid exponentially blows up ( leading to nugget formation \cite{Afshordi:2005ym}) whereas normal radiation perturbation oscillates and standard dark matter perturbation grows much slower than neutrino-scalar fluid.}
    \label{fig:pertubations}
\end{figure}

\section { Solving for the static Profile of the Dark Matter Nugget} \label{sec:static}
The formation of the dense fermionic object in the presence of a scalar mediated force (eg. soliton star, fermionic Q-star) has long been an interesting research topic in different astrophysical contexts \citep{Lee:1986tr, Lynn:1989xb}.  The formation of a  relativistic star in chameleon theories  has also drawn recent interests \citep{Upadhye:2009kt}. In the context of neutrino dark energy, formation of neutrino nugget at very late time ($z$ $\sim $ few  ) has been studied in \cite{Afshordi:2005ym}, while large (\rm{Mpc}) stable structures of neutrinos clump in the presence of an attractive quintessence scalar has been discussed in  Ref.~\cite{Brouzakis:2005cj, Brouzakis:2007aq,Pettorino:2010bv,Wintergerst:2009fh}. As the quintessence field is extremely light and comparable to Hubble expansion rate ($m \sim H $), the range of the fifth force is very large and a neutrino clump can easily be of the order of Mpc size \citep{Casas:2016duf,Brouzakis:2007aq}. In all of these studies, these Mpc sizes neutrino clumps form at a very late time in cosmic evolution (around $z \sim $ few) and thus contribute only a tiny fraction to the dark matter relic density. Whereas we focus on a scalar field with much heavier mass ($m_{\phi} \sim 10^{-3}$ \rm{eV})  and  the formation of nuggets takes place at a much earlier redshift ($z_F \gtrsim 10^6$). We show that, as a result,  these compact nuggets can, in principle, comprise the entire cold dark matter relic abundance.
 
In this section we  first discuss the qualitative features of the nugget formation, and then numerically solve the bubble profile  for the static configuration. For a detailed  analysis of the nugget collapse process involving non-linear dynamics, see \cite{Narain:2006kx}. 
We assume that for a critical over density $\delta$, the sound speed of perturbation becomes negative and the fifth force attracts neutrinos within a Compton sphere of the scalar and finally, Fermi pressure intervenes and balances the attractive  scalar force and stable nugget profile is achieved.
One can in principle solve the static configuration of nugget by solving coupled Klein-Gordon equation. Taking the metric to have the form 
\begin{equation} 
ds^2= -A(r) dt^2 + B(r) dx^2 + r^2 d\Omega^2
\end{equation}
 The gravitational mass of the system can be found from asymptotic form of $A(r)$ for $r \rightarrow \infty $ and is given by \cite{Brouzakis:2005cj}

\begin{equation}
M(r)= \int 4 \pi r^2 dr \left[\frac{1}{2} \left(\frac{d\phi}{dr}\right)^2 + V(\phi(r)) + \rho_{\psi_{1}} (r)\right]
\end{equation} 
 where, the first term is the gradient energy and the other terms correspond to the  $\phi$ dependent fermion mass and the scalar potential $V(\phi)$.
 
For our case, the size of the nuggets is very tiny and the scalar force is much stronger than gravity; thus, we can safely ignore gravity and will work with the Minkowski metric.

 A general feature of the scalar field static configuration is that the scalar \textit{vev} changes as a function of distance from the center of the nugget and takes an asymptotic value far away from it. The fermion number density, Fermi pressure are all functions of position through $\phi(r)$.
 
 As the fermion mass also varies radially, we adopt the Thomas-Fermi approximation as done in \cite{Brouzakis:2005cj} to find the static configuration. In the Thomas-Fermi approximation, one assumes, at each point in space there exists a Fermi sea with local Fermi momentum $p_F (r)$. With an appropriate scalar potential, it has been shown in  \cite{Lee:1986tr,Lynn:1989xb} that a degenerate fermionic gas can be trapped in compact objects (even in the absence of gravity) which has been called as soliton star or fermion Q-star. These analyses were done with the zero temperature approximation and in this case, fermions can be modeled as ordinary nuclei. To get an estimate of the nugget radius, we will also work with zero temperature approximation. Soon, we will see that  our numerical solution has similarities with these previous works where fermions are extremely light inside the nugget and become increasingly more massive as one moves towards the nugget wall. This ensures positive binding energy and thus the stability of the nuggets.

 \subsection{Static Solution}

 Static solutions of this type are mainly governed by two equations. We refer to \cite{Brouzakis:2005cj,Lee:1986tr,Chanda:2017coy} for detailed derivation of these equation. Briefly, the first one is the Klein-Gordon equation for $\phi(r)$ under the potential $V(\phi)= m^2\phi^{2}$ where the fermions act as a source term for $\phi(r)$. The other equation tells us how the attractive fifth force is balanced by local Fermi pressure. 

As we will be working in the weak field limit of general relativity, the  equations can be expressed as 
\begin{equation}\label{StaticPhi}
\phi'' + \frac{2}{r} \phi' = \frac{dV(\phi)}{d\phi} - \frac{d\mathrm{ln}[m(\phi)]}{d\phi} T_{\mu}^{\mu}. 
\end{equation}

\begin{equation}\label{FermiMomentum}
\frac{dp}{d\phi}= \frac{d[\mathrm{ln}(m(\phi))]}{d\phi} ( 3 p -\rho).
\end{equation}

Then, \eqref{StaticPhi} can be rewritten in the simpler following form valid inside the bubble \citep{Brouzakis:2005cj}, 
\begin{equation}\label{eq:StaticSolPhi}
\phi'' + \frac{2}{r} \phi' =-\frac{d(p-V(\phi))}{d\phi}. 
\end{equation}

Outside, there is no pressure, thus this simplifies further
\begin{equation}
\phi'' + \frac{2}{r} \phi' =\frac{dV(\phi)}{d\phi}.
\end{equation}  

Now with the Thomas Fermi approximation, 
the distribution function is given by \cite{Brouzakis:2005cj}, 
\begin{equation}
f(p)=\left(1+e^{\frac{\sqrt{p^2(\phi) + m^2(\phi)}-\mu(r)}{T(r)}}\right)^{-1}
\end{equation}

 With the zero temperature assumption, the distribution function becomes a step function and pressure, energy density and number density
take the form:

\begin{equation}\label{eq:pFormula}
p(r)=\frac{1}{4 \pi^3} \int^{p_{F}(r)} d^3p\frac{p^2}{3 \sqrt{p^2 + m(\phi)^2}},
\end{equation}

\begin{equation}\label{RhoRelFormula}
\rho(r)=\frac{1}{4 \pi^3} \int^{p_{F}(r)} d^3p \sqrt{p^2 + m(\phi)^2},
\end{equation}

\begin{equation}\label{eq:nFormula}
n=\frac{1}{4 \pi^3} \int^{p_{F}(r)} d^3p \times 1 = \frac{p_F^3}{3 \pi^2}.
\end{equation}

Thus there are two unknown quantities, namely $\phi(r)$ and $p_F(r)$, whose values are determined by solving the two coupled equations \eqref{FermiMomentum} and \eqref{eq:StaticSolPhi}.	Evaluating  the integrations in eq. \eqref{eq:pFormula}-\eqref{eq:nFormula}, we find an explicit form for the trace of energy-momentum tensor, $T_{\mu}^{\mu}=(\rho - 3 p)$:
\begin{align}\label{eq:EnergyMomentum}
T_{\mu}^{\mu}&=\frac{m^2}{2 \pi^2}\left(p_F \sqrt{p_F^2 + m^2} - m^2 \mathrm{ln}\left(\frac{p_F + \sqrt{p_F^2 + m^2}}{m}\right)\right).
\end{align}

Furthermore the pressure is found to be of the form
\begin{equation}\label{eq:p}
p=\frac{m^{2}}{8\pi^{2}}\left[\frac{2p_{F}^{3}}{3m^{2}}\sqrt{p_F^2 + m^2}  - \left(p_F \epsilon_F - m^2 \mathrm{ln}\left(\frac{p_F + \epsilon_F}{m}\right)\right) \right],
\end{equation}
where $\epsilon_F^2 = p_F^2 + m^2 $. 
\begin{figure}
\centering
\includegraphics[scale=0.47]{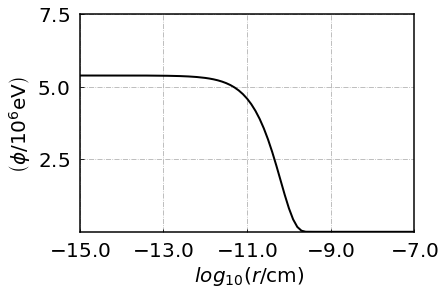}~~~
\includegraphics[scale=0.47]{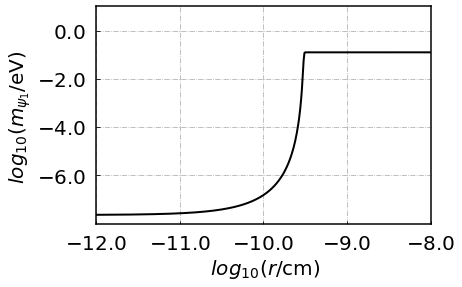}
\includegraphics[scale=0.47]{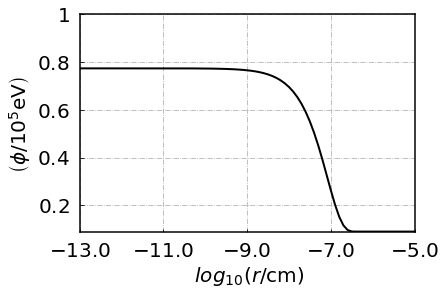}
\includegraphics[scale=0.47]{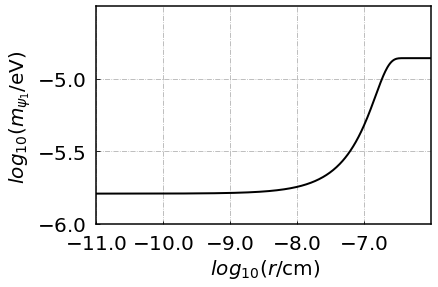}
\caption{\textit{Left}: Numerical solution for Static configuration of the scalar field coupled to light singlet fermion. We see that inside the nugget the scalar vev is higher and outside it merges asymptotically to its cosmological value. \textit{Right}: Variation of trapped fermion mass inside the nugget. This is obtained from the static profile of scalar field inside the nugget. The plots in the above and below corresponds to $z\sim 8.56\times 10^{4}$ and $z\sim 5.8\times 10^{7}$ respectively.}\label{StaticProfile}
\end{figure}

\subsection{Initial conditions and model parameters}
Using eqs. \eqref{eq:p} and \eqref{eq:EnergyMomentum} in eq. \eqref{FermiMomentum}, we obtain an equation for  the local Fermi momentum $p_F(r)$ which needs to be solved with proper boundary conditions. This initial condition $p_F^0$ is chosen appropriately to match the total number of fermions initially within a Compton volume of the scalar before the onset of instability. First, we solve for $p_F(r)$ and using that we numerically solve the 
Klein-Gordon equation (5) to obtain the  $\phi(r)$ profile.

 The Klein-Gordon equation  closely resembles the dynamics of a particle moving under a Newtonian potential $p-V$ when ``$r$'' is replaced by time ``$t$'' and ``$\phi$'' is replaced by the position of the particle. The static solution corresponds to a final particle position which is  asymptotically at rest. This is only possible for a particular set of initial condition $\phi(0)\equiv\phi_0$ and $\phi^{\prime}_0$
For the solution to be well behaved at the center of the nugget, $\phi$ should not have a term like $r^{-n}$ which demands
$ \phi^{\prime}_{0}= 0 $. The other initial condition is obtained by numerical iteration i.e. by identifying a particular $\phi_{0}$ for which the numerical solution gives $\phi(\infty)= \phi_{\rm cosmo}$, where
 $\phi_{\rm cosmo}$ is the cosmological value outside the nugget. For our case, $\phi_{\rm cosmo}=0$ which is the minima of the $m_\phi^2 \phi^2 $ potential. 
Once we find $p_F(r)$ and $\phi(r)$, it is straightforward to obtain the fermion mass $m_{\psi_{1}}(r)$ and fermion number density $n(r)$ as a function of distance from the center of the nugget. The radius of the nugget is determined when the number density drops to zero.

Next, we numerically find the static profiles for the scalar field for two toy examples to show that indeed stable nugget can be formed. 
\subsubsection{Example 1}
We first consider fermions of mass $m_{\psi_{1}}$ which are trapped in nuggets and using, $m_{D}^{2}/\tilde{\lambda} = 0.125~{\rm eV}^{2}$, so to obtain an example static profile for $m_{\psi_{1}} = m_{D}^{2}/\left(\tilde{\lambda}\phi_{\rm static}\right)\simeq 2.4\times 10^{-8}~ {\rm eV} $. The Fermi momentum $p_{F}(r)$ is fixed by identifying a value for which $\phi(\infty) \sim \phi_{\rm cosmo} =0$, one solution is:
  \begin{equation}
{p_{F}(0) \Big|_{\phi_{0}} = 5.8 \times 10^{3} ~{\rm eV}.}
  \end{equation}

   The inferred $p_{F}$ is then used on the energy-momentum tensor $T^{\mu}_{\mu}$ to solve eq. \eqref{StaticPhi}. We take $m_{\phi} = 1.34\times 10^{-3}~{\rm eV}$ with the external potential $V(\phi) = m_{\phi}^{2}\phi^{2}$ and the resulting static profile of the scalar field $\phi$  (as shown in Fig.\ref{StaticProfile}), is obtained from the  boundary condition 
   \begin{equation}
{\phi[3.2\times 10^{-10}~{\rm cm}] =1~ {\rm eV}}
   \end{equation}
   and we evaluate the profile for  $3.2\times 10^{-10}~{\rm cm}<r<1.98\times 10^{-5} ~{\rm cm}$.
   
    The mass variation of trapped fermion inside the nugget is depicted as in Fig.3. The total nugget mass is found to be $M_{\rm nug}\simeq 2.54\times 10^{6}~{\rm eV}$, which with the chosen scalar field mass, corresponds to a dark matter relic density at the nugget formation of order $\rho_{F}^{{\rm DM}} = 6.13\times 10^{3}~{\rm eV}^{4}$.

\subsubsection{Example 2}
We repeat the same treatment with a different choice for the Fermi momentum boundary value
 \begin{equation}
{p_{F}(0) \Big|_{\phi_{0}} = 4.6 \times 10^{5} ~{\rm eV},}
  \end{equation}
and boundary condition for the $\phi$ solution
   \begin{equation}
{\phi[4.2\times 10^{-7}~{\rm cm}] =9\times 10^{3}~ {\rm eV}}
   \end{equation}
    so to obtain an static profile for $4.2\times 10^{-7}~{\rm cm}<r<7.9\times 10^{-5} ~{\rm cm}$., with $m_{\psi_{1}} = m_{D}^{2}/\left(\tilde{\lambda}\phi_{\rm static}\right)\simeq 1.6\times 10^{-6}~ {\rm eV} $ and $m_{\phi} = 2.68\times 10^{-6}~{\rm eV}$. This parameter set suggests a total nugget mass $M_{\rm nug}\simeq 9.8\times 10^{28}~{\rm eV}$ and a dark matter relic density of $\rho_{F}^{{\rm DM}} = 1.89\times 10^{12}~{\rm eV}^{4}$ at the time of nugget formation.

\subsection{Possibility of a homogeneous marginally stable tenuous phase of neutrino gas? }

  Now we would like to explain an important issue regarding our zero temperature approximation of nugget formation. Here we discuss whether our assumption of forming cold nuggets surrounded by vacuum as a result of phase transition  is realistic or not. Because another possible outcome can be formation of  dense nuggets, surrounded by a tenuous hot gas of neutrinos. But it seems like instability in mass varying neutrino  prefers the former!
 The explanation for this  can be   obtained  by following the discussion in the paper \cite{Afshordi:2005ym}  where a detailed  instability study in "mass varying of neutrino dark energy” was first  carried out.  They showed that the density of neutrinos in the  tenuous gas phase (which is not trapped  in nugget) is exponentially suppressed  and that the most probable outcome of the instability is the formation of dense nuggets  surrounded by practically empty space without any marginally stable tenuous gas of neutrino interacting with scalar.
The main reason they found  why we don’t see a tenuous gas phase in this phase transition  is mainly for two reasons . First of all, the time scale of instability  $ ( m_{\phi})^{-1} << H^{-1} $  is much smaller than Hubble time scale unlike the standard quintessence model of dark energy . As mass of the scalar field here is much  higher than Hubble parameter during MRE,  one can assume a very rapid formation of nugget compared to cosmological Hubble time scale. They showed that  the imaginary speed of sound is indeed scale-independent in any non-relativistic MaVaNs models, making the instability most severe at the microscopic scales  ($\simeq  m_{\phi}^{-1}$).  The fraction of neutrinos in the gas phase is computed by the equilibrium of evaporation rate from and accretion rate onto the surface of the nuggets. However, the large density contrast between inside and outside of the nuggets implies that only neutrinos with a large Lorentz factor can escape the nuggets. This is also obvious from our plot of variation of neutrino mass as function of distance from nugget center. 
So with the above  arguments we believe our zero temperature assumption will be the most probable scenario of such phase transition  but of course non-linearity can bring surprise and is kept for future research.

\section{Cosmology of phase transition} \label{sec:cosmology}

Dark Matter nugget formation from instability in neutrino is a unique phenomenon and it may have some interesting effect on both CMB and linear structure formation. In this section, we explore these effects if we let this phenomenon to account for the entire DM density in the universe. To find the effect of the nugget formation in CMB and linear structure formation, we need to solve for the linear as well as perturbation equations for the whole cosmic history. We adopt the generalized dark matter(GDM) formalism describe our interacting $\phi$-$\psi_1$ fluid. In this formalism, the background equation of the fluid is parameterized by its equation of state, $w_{\phi-\psi_1}$ which we take to be a function of the redshift $z$. At early times, since we assume the fluid to be relativistic, the coupling between neutrino and the scalar field can be ignored. So, we take $w_{\phi-\psi_1}\sim\frac{1}{3}$ at high redshifts. Then, as the neutrinos become non-relativistic around a critical redshift $z_F$, an EDE like phase appears when the scalar field relaxes into the effective minima followed by an intermediate transition period where the effective equation of state goes below zero to $w_{\phi-\psi_1} \sim -\frac{1}{3}$. During this period, sound speed square, $c_s^2$ also becomes negative and we have instability in the fluid. Immediately after that, phase transition happens as a result of the instability and the fluid transitions into DM nugget state, where $w_{\phi-\psi_1}=0$. 

The perturbations equations for our fluid in synchronous gauge from the GDM formalism are given by:
\begin{eqnarray}
\dot{\delta} &=& -(1+w)\left(\theta+\frac{\dot{h}}{2}\right)-3(c_s^2-w)H\delta \label{eq:delta_dot}\\
\dot{\theta} &=& -(1-3c_s^2)H\theta + \frac{c_s^2k^2}{1+w}\delta
\end{eqnarray}

\begin{figure}
\centering
    \includegraphics[scale=0.55]{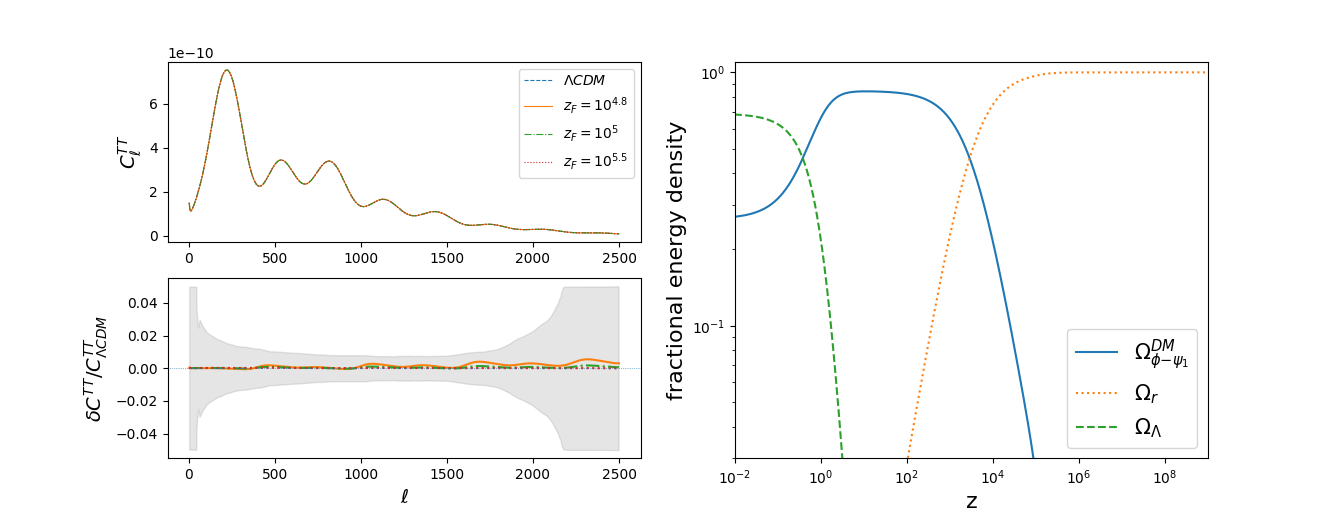}
    
    \caption{\textit{Left}: CMB TT spectrum for three transition redshifts $10^{4.8},10^5,10^{5.5}$ along with \LCDM. The residuals from Planck \LCDM data can be seen below. The grey area in the background is the error bar set by Planck data. One can see as the phase transition epoch gets closer to CMB, the deviation increases as expected. \textit{Right}: Evolution of the fractional energy density of radiation(photon+standard neutrinos), neutrino-scalar fluid and $\Lambda$ through entire cosmology for $z_F = 10^5$.}
    \label{fig:CMB}
\end{figure}

To solve the background and the perturbation equations, we modify the CLASS Boltzmann code \citep{Blas:2011rf} to replace the CDM component with an extra fluid component which describes our fluid through $w_{\phi-\psi_1}$ and $c_s^2$. In Fig.\ref{fig:pertubations}, we indeed see that around $z_F$, the density fluctuations of the $\phi$-$\psi_1$ fluid shoots up to a huge value compared to those of the other components such CDM and neutrinos from standard \LCDM  scenario. In Fig.~\ref{fig:CMB}(\textit{right}), we plot the fractional energy density of the dominant components of the universe at different times against redshift $z$ for the case $z_F=10^5$. We see that in early times, radiation is the dominant component as usual but a tiny yet significant portion of the universe ($\sim10^{-2}$) consists of $\phi$-$\psi_1$ as dark radiation. Then around $z_F$, $\phi$-$\psi_1$ transitions from radiation into matter like state and $\Omega_r$ starts to fall while $\Omega_{\phi-\psi_1}$ rises. The Matter-radiation equality happens at $z\sim3400$. The DM nugget energy density dominates the universe until $\Omega_\Lambda$ crosses $\Omega_{\phi-\psi_1}$ at late times ($z\sim10^{-1}$). The value of $\Omega_{\phi-\psi_1}$ today comes out to be $\sim0.2637$ which is equal to the dark matter density today according to $\Lambda$CDM.

In Fig.\ref{fig:CMB}(\textit{left}), we plot the CMB temperature anisotropy spectrum along with the residuals from Planck \LCDM  data \cite{Ade:2015xua} for three different critical redshifts,  $z_F=10^5,10^{5.5},10^{4.8}$. We see that the later the critical redshift is, the more it deviates from \LCDM. For the case $z_F=10^{4.8}$, the residuals go beyond the error bar set by Planck. The residual plot is also dependent on the extent of the transition period which we haven't considered in this paper. A proper MCMC analysis is needed to find a suitable $z_F$ and the extent of the transition period which is left for future work.


\subsection{DM density and $\Delta N_{\textrm{eff}}$}

One can easily estimate dark matter density as nuggets are highly non-relativistic and assuming  one nugget forming  in each Compton volume  of the scalar  $\sim (m_{\phi}^{-1})^3$, the energy density of dark matter at the formation redshift $z_F$ is simply given by
\begin{equation}
\rho_{F}^{DM}= \frac{M_{\rm nug}}{\frac{4}{3} \pi (m_{\phi}^{-1})^3}.
\end{equation}

Evolving this to today implies  
$ \rho_{DM}^0 = \rho_{\rm DM,F} (1+ z_F)^{-3}$ where  $M_{\rm{nug}}$ is the nugget mass. We mainly focus on the case when nuggets form only a sub-fraction of DM and most of the energy density of EDE phase goes into scalar dynamics.  
 At early time $z > 10^{5}$  before neutrino turns semi-relativistic and starts to couple to scalar field,  the neutrino-scalar fluid was behaving like dark radiation. But during CMB this fluid does not  does not  behave as radiation as phase transition has already changed its equation of state during MRE. But never the less very small scale modes which was inside the horizon at very early epoch $ l > 200 $  would be slightly affected and have tiny imprint on CMB too. To see this effect, we numerically solve and plot the CMB power spectra deviation in Fig.~\ref{fig:CMB} and we see that a redshift of transition $z_F \geq 10^5$ is within the error bar of Planck.
 Also, in our MCMC analysis in next section, we will see that statistically our model performs quite well when confronted with Planck, local $H_0$ and BAO data sets.\\
Never the less,  if one is interested in very early epoch  and especially epoch of  BBN, the amount of dark radiation in our model is estimated \cite{Das:2020nwc}
\begin{equation}
 \label{eq:y:3}
  \Delta N_{\rm eff} \approx \frac{\rho_{\rm CDM}(0)}{\rho_{\phi-\psi_1}(0)} \approx 0.2 \left( \frac{\Omega_{\rm CDM} h^2}{0.1199} \right) \left ({\frac{10^5}{1+z_F}} \right ).
\end{equation}
This excess radiation is well with in the BBN bound.  The matter power spectra  form this late dark matter neutrino nuggets has sharp cut-off in small scales which has lot of interesting small scale effects and has recently been studied in \cite{Das:2020nwc}.

\subsection{Nugget life time and production of scalar radiation}
 \label{subsec:constrains}

 In our model, after its formation, nuggets can produce scalar dark radiation through two body annihilation of  sterile neutrino like particle in the dense environment inside the nugget. In this section, we  show that the nuggets  are stable enough to be dark matter by calculating the annihilation rate of $\psi_{1}$  into $\phi$ inside the nugget. As the scalar is much lighter ($10^{-5}$ \rm{eV}) than fermion, we are interested in the annihilation process: \, \, $  \psi_1 + \psi_1 \rightarrow \phi + \phi $. 

From our numerical results, it is clear that  fermion mass $m_{\psi_{1}}$ is extremely light closer to the center of the nugget and becomes heavier at the wall. For the same reason, integrating out heavy field is a valid approximation inside the nugget (numerically verified) and the fermion mass term is given by $ \left(m_D^2/f(\phi)\right) \psi_{1} \bar{\psi_{1}}$.   
Inside the nugget, the coupling  constant $\kappa$, between light fermion and $\phi$ is defined through the following interaction term

 \begin{align}
 \kappa_{\rm in}\delta \phi \bar{\psi_{1}} \psi_{1}\sim\frac{dm_{\psi_{1}}}{d\phi}\Big\vert_{_{\phi_{\rm static}}} \delta \phi \bar{\psi_{1}} \psi_{1},
 \end{align}
 where $\delta \phi$ is fluctuation in the $\phi$ field.

We find that $\kappa_{\rm in}$, the value of the coupling inside the nugget, is very tiny: $\kappa_{\rm in} \sim 10^{-14}$. 
The nugget lifetime  can be estimated for a given $\kappa$. In terms of the number density of the fermion inside the nugget($n$), the annihilation rate is given by
 \begin{align}
 \frac{dn}{dt} \sim  \frac{n^{2} \kappa^{4} }{32 \pi E^{2}_{\rm CM}} 
 \end{align}
 
 where $E_{\rm CM}\sim\sqrt{p_{F}^{2}+m_{\psi_{1}}^{2}}$ is the center of mass energy. Integrating, we can estimate the half life of the nugget
 \begin{align}
 \Delta t_{1/2} \sim \frac{n_0}{V \Gamma_0} \sim \frac{1}{n_{0}\frac{\kappa_{\rm in}^{4}}{32\pi E_{\rm CM}^{2}}} ,
\end{align}
where $V$ is the volume of the nugget and $n_{0}$ is the central number density at $r=0$. Taking the specific example of a dark matter nugget studied in the previous example, and substituting the values from this
 numerical solution, we find  half life of the nugget is roughly  $ \Delta t_{1/2} \sim 10^{44} $ s. This is much larger than the age of the universe $\sim 10^{17}$ sec.

\begin{figure}[ht]
\centering
\begin{tikzpicture}[node distance=2cm and 2.5cm]
\coordinate[label = left:$\psi$](e1);
\coordinate[below =3 cm of e1, label = left:$\psi$](e2);
\coordinate[right =4 cm of e1, label = above:$-i\lambda$](e3);
\coordinate[right =4 cm of e2, label = below:$-i\lambda$](e4);
\coordinate[right =8 cm of e1, label = right:$\phi$](e5);
\coordinate[right =8 cm of e2, label = right:$\phi$](e6);
\draw[fermion] (e1) -- node[] {} (e3);
\draw[fermion] (e2) -- node[] {} (e4);
\draw[scalar] (e3) -- node[] {} (e5);
\draw[scalar] (e4) -- node[] {} (e6);
\draw[fermion] (e3) -- node[label = right:$\psi$] {} (e4);
\end{tikzpicture}

\caption{Dark matter to $\phi$ field production inside the nugget.}
\end{figure}
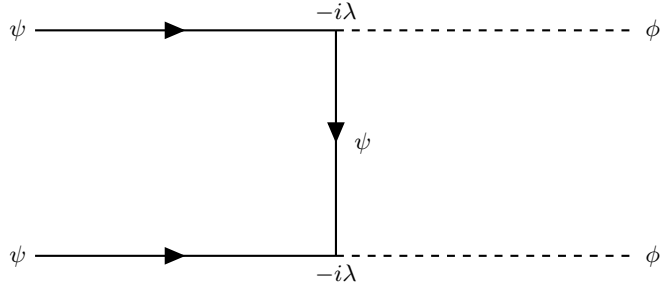

\section{ Hubble anomaly in connection to our model}
\label{sec:hubble}
 
 In this section, we discuss the recent Hubble anomaly in the context of our model mainly for two reasons.
The first reason is- before the nugget formation, the neutrino-scalar fluid naturally goes through short early dark energy domination as explained before. Given that the early DE model has been proposed as one most prominent solution of Hubble anomaly, we explore if our model can be realized as a successful candidate for EDE scenario arising from (sterile) neutrino-scalar interaction.\\
 
 The second reason is, in a generic EDE model, from the theoretical perspective, the physics of matter-radiation equality and early dark energy are completely disconnected - which requires fine-tuning  in order for them to appear nearly simultaneously. 
 Our model is based on physics where scalar mass is of the order of neutrino mass scale (unlike in \cite{Sakstein:2019fmf}, where scalar mass is tuned to an extraordinarily small value, like in the case of quintessence theories) as proposed originally in the theories of neutrino dark energy and the similarity of neutrino mass and MRE scale naturally gives early dark energy phase prior to MRE. In a neutrino-scalar interaction scenario, the background equation for the neutrinos is given by 
\begin{equation}
\dot{\rho}_{\psi_1}+3H(\rho_{\psi_1}+p_{\psi_1}) = \frac{dln(m_{\psi_1})}{d\phi}\dot{\phi}(\rho_{\psi_1}-3p_{\psi_1}) \label{eq:bg_nu}
\end{equation}
where the right-hand side is the interaction term. We see that when neutrinos are relativistic, the interaction term is negligible since $(\rho_{\psi_1}-3p_{\psi_1})\sim0$. Therefore, the interaction is only prominent when the neutrinos become non-relativistic around $z_F$ and its equation of state drops below $ w < -1/3$ ( the exact value depends on the potential) and the fluid starts to behave like dark energy. This early phase of DE is only active for a short duration(see right panel of Fig.~\ref{fractional energy}) followed by its disappearance as nugget formation takes place.
 The excess dark energy gives a local boost to local Hubble expansion rate and  the sound horizon for acoustic waves in the photon-baryon fluid, $r_s$ is given by
\begin{equation}
    r_s = \int_{z_{eq}}^\infty \frac{c_s}{H(z)}dz.
\end{equation}
Since H(z) in the EDE scenario was higher than that of $\Lambda$CDM for a short period of time, it would mean that $r_s$ is reduced compared to the $\Lambda$CDM model. The angular diameter distance to the last scattering surface is given by
\begin{equation}
    d_A = \frac{r_s}{\theta_s},
\end{equation}
where $\theta_s$ is the angular scale of the sound horizon at matter-radiation equality. This angular scale is fixed as it is solely determined by CMB observation and hence its value is not sensitive to any model. This implies that $d_A$ also gets reduced in this scenario. Since $d_A$ is inversely proportional to the Hubble constant, this means $H_0$ is higher than it is in $\Lambda$CDM.\\

\begin{figure}
    \centering
    \includegraphics[scale=0.28]{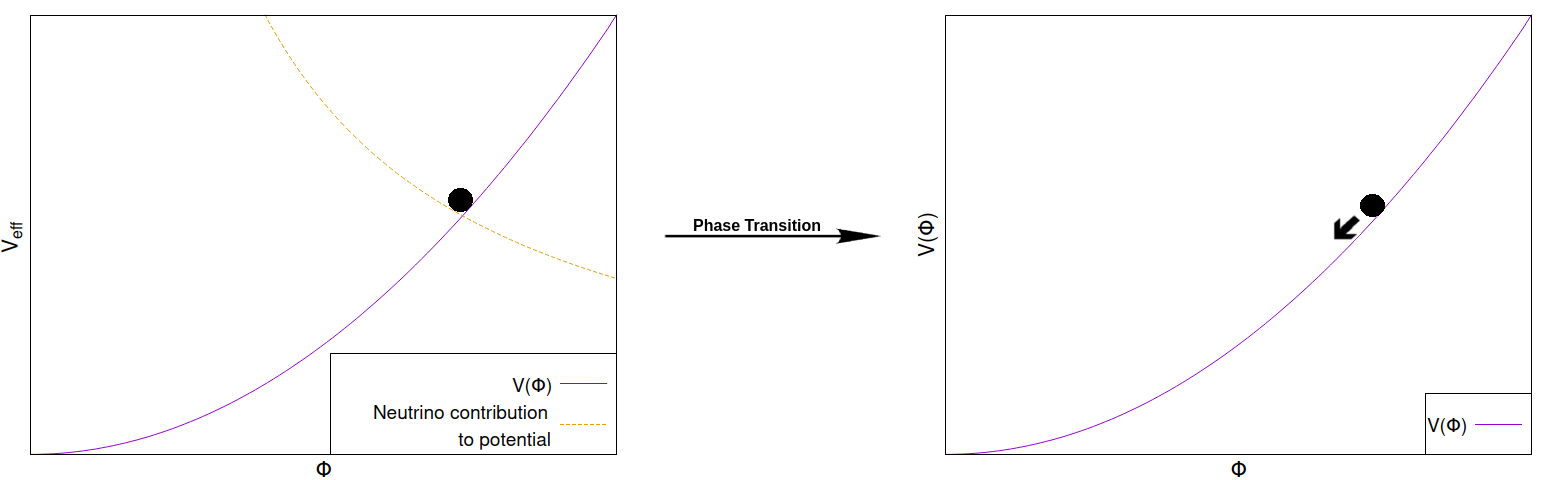}
    \caption{Pictorial demonstration how the effective minima controls the field dynamics before nugget formation and behaves like early DE. As soon as nugget forms, the support from $\phi $ dependent neutrino density vanishes and field starts to roll of fast along the potential.  }
    \label{fig:veff}
\end{figure}

\subsection{ Simple model \textbf{(A)}: Neutrino EDE entirely  transitions to CDM }
First we consider the most simple case when 
the entire energy of the EDE phase goes into the dark matter state. This is assuming all the neutrinos clumps inside the nugget and the scalar field after being released coherently  oscillates around quadratic potential which  also behaves like dark matter provided it does not decay into other lighter particles \cite{Bjaelde:2010vt}. The evolution of equation of state of the scalar fluid is shown in right panel of Fig.~\ref{fractional energy} where in most of the radiation dominated era ( except when neutrino turns non-relativistic)  the neutrino-scalar fluid behaves like radiation with $w= 1/3$ followed by a dark energy like behavior $w \leq - 1/3$ ( the exact value depends on potential $V(\phi)$ and finally after nugget formation the equation of state becomes likex CDM $w=0$.\\ 
We numerically find that this case cannot (or only partially) resolve  the Hubble anomaly. This is expected as from original EDE study \cite{Poulin:2018cxd} where it is shown that the early dark energy phase should dilute at least like radiation or faster to resolve Hubble anomaly completely. However, now we discuss that in extended models of our scenario, it is highly possible that a major fraction of early DE phase indeed goes into the scalar field which redshifts much faster than radiation. This is realised  when the quadratic potential is slightly modified at $\phi =0$  and the scalar field redshifts  away its initial energy  through a kinetic energy dominated phase, similar to \cite{Alexander:2019rsc, Lin:2019qug}.

\begin{figure}
\centering
    \includegraphics[scale=0.4]{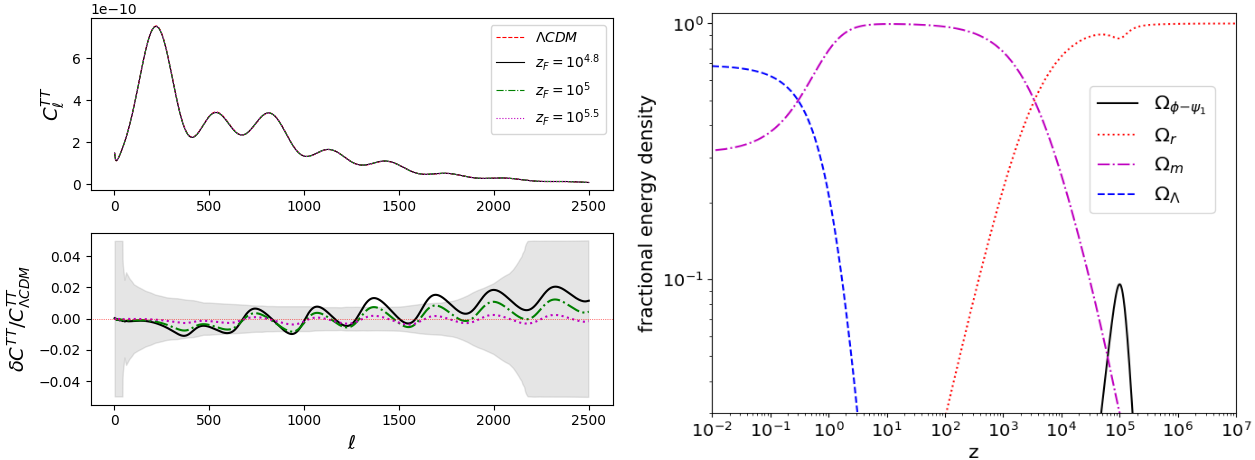}
    
    \caption{Plots for model \textbf{(B)} with $V(\phi)\sim log(\phi)$. \textit{Left}: CMB TT spectrum using for three transition redshifts $10^{4.8},10^5,10^{5.5}$ along with \LCDM. The residuals from Planck \LCDM data can be seen below. The grey area in the background is the error bar set by Planck data. One can see as the phase transition epoch gets closer to CMB, the deviation increases as expected. \textit{Right}: Evolution of the fractional energy density of radiation(photon+standard neutrinos), neutrino-scalar fluid and $\Lambda$ through entire cosmology for $z_F = 10^5$.}
    \label{fig:CMB1}
\end{figure}

\begin{figure}
\centering
    \includegraphics[scale=0.4]{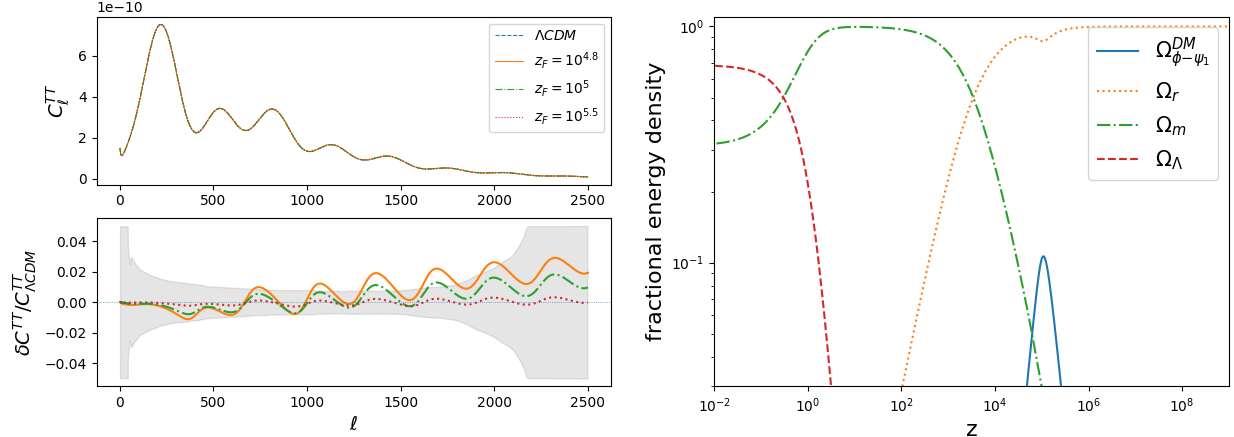}
    
    \caption{Same as Fig.~\ref{fig:CMB1} with $V(\phi)\sim\phi^2$.}
    
\end{figure}

\pagebreak \subsection{ Extended model \textbf{(B)}: Neutrino EDE dilutes as Kinetic energy dominated phase \textit{Kination} }

\subsubsection{ Evolution of scalar  immediately after nugget formation}

As discussed before, most probable outcome of the instability is the formation of dense non-linear structures of nuggets, surrounded by practically empty space where the scalar starts to role toward its true minima as the effective minima  disappears in a time scale which is much smaller than  the local Hubble time. Thus the dynamics of the fluid is indeed dominated by a kinetic energy phase and EDE energy dilutes much faster than radiation. 
After the nugget formations, neutrinos in the “gas phase ( a tiny fraction of total neutrino)” continuously lose energy to Hubble expansion, and nuggets, due to their small volume fraction, cannot significantly affect the gas phase. As a result most of the neutrinos should accrete into the nuggets. In fact,  in Sec. V of \cite{Afshordi:2005ym}  they showed that within the assumption of
thermal equilibrium, this process instantly exhausts the
neutrinos outside the nuggets and scalar field $\phi$ settles at its true vacuum within the gas phase.\\
In the conventional quintessence models, similar to inflationary models, the scalar field is slowly rolling, and therefore its effective mass is smaller than the Hubble expansion rate. In contrast, in mass varying dark energy model, the scalar field sits at the instantaneous minimum of its potential, and the cosmic expansion only modulates this minimum through changes in the local neutrino density. So once the nugget forms surrounded by practically empty space, the scalar field  quickly settles to its true minima globally. As a result our kination picture is the most natural outcome of this scenario.

So as stated before, unlike the general EDE model, our early dark energy phase is not controlled by local Hubble friction rather neutrino density. But after the nugget formation, scalar decouples from neutrino density and its evolution is determined  by Klein Gordon equation 

\begin{equation}
\ddot{\phi} + 3 H \dot{\phi} + V(\phi)=0
\end{equation}

It is instructive to note that  here we have neglected the $\nabla^2 \phi$ term. In our case, though scalar field profile varies around nuggets, we realise that nugget size ( as seen from Fig.2 ) is much smaller than the Compton length of the scalar field. As can be seen from the same figure - the scalar field value varies within a length scale $ \leq 10^{-10} $ cm where the the Compton length of the scalar $m_{\phi}^{-1 } \simeq  10^{-2}$cm. ( this corresponds to our scalar mass $10^{-3} \rm{eV}$ in the first numerical example). So within our assumption, the scalar field would sense the nuggets as point objects ( as the spatial variation scale of $\phi$ is $10^{-8}$ times smaller than scalar Compton length). On top,  the number density of such point like nugget within one Compton volume is extremely low-- more or less one nugget within each Compton volume of scalar. This also justifies our assumption.
Our situation is similar to chameleon theory \cite{Khoury:2003aq} where in small distance (solar to galactic scale) chameleon scalar spatial variation is non-negligible but in  cosmological distance ($\geq \rm{Mpc}$ ) the field is controlled by Klein Gordon equation without a $\nabla^2 \phi$ term.  But in our scenario, if  nuggets  merge and continuous merger make the final nugget radius comparable to scalar Compton length, the $\nabla^2 \phi$ contribution might be important. This will be clear when we study nugget collision and change of scalar profile due to merger. This is beyond the scope of present work and has been kept for future research. Specially, if larger sterile neutrino halo forms due to merger as well as  virialisation, one needs to incorporate $\nabla^2 \phi$ contribution to K.G. equation and it would be interesting to see how it changes scalar field dynamics. It is also important to note that  the initial condition for our case  when the field starts to be dynamical would be very different than \cite{Poulin:2018cxd} where field starts to roll of when mass of the field becomes larger than Hubble friction. Moreover, it is possible to have a steeper potential where the scalar field is held in the effective minima and rolls off very fast (dominated by kinetic energy \citep{Alexander:2019rsc,Braglia:2020bym}) as soon as the field is released during the onset of nugget formation. In this scenario it is shown that the EDE  rapidly redshifts away $\sim 1/a^6$ and thus solving the Hubble anomaly.

\subsubsection{ Choice of potential }
We have pointed out earlier that a quadratic potential is most natural in mass varying neutrino DE in the particle physics context of re-normalization. On the other hand, a quadratic potential was recently used  in the Acoustic Dark Energy (ADE) model of \cite{Lin:2019qug} where $ V(\phi)\sim \phi^2 $ for $\phi > 0$ and $ V(\phi)= 0 $ for $\phi \leq 0$. Here, the whole potential energy converts into kinetic energy and hence $w = P/\rho = (\frac{1}{2}m\dot{\phi}^2-V(\phi))/(\frac{1}{2}m\dot{\phi}^2+V(\phi)) = 1$. $V(\phi)$ is cut off for negative values of $\phi$ such that there is no oscillation and no kinetic energy can convert back into potential energy. Another potential where it is possible realize similar scenario by taking a potential which is within the frame work of  $\alpha$-attractor potential where inflationary potential is unified with late DE like behavior. In the case "C" of \cite{Braglia:2020bym}, they take a potential $V(\phi) \sim \frac{(1+\beta)^4tanh^8(\phi/\sqrt{6\alpha}M_{pl})}{[1+\beta tanh^8(\phi/\sqrt{6\alpha}M_{pl})]^4}$. This potential is flat enough(see Fig.1 of \cite{Braglia:2020bym}) so that the scalar field does not oscillate around the minimum after rolling off and instead, it remains in kination until its fractional energy density becomes insignificant and then freezes out after exhausting all its inertia.

\begin{figure}
    \centering
    \includegraphics[scale=0.54]{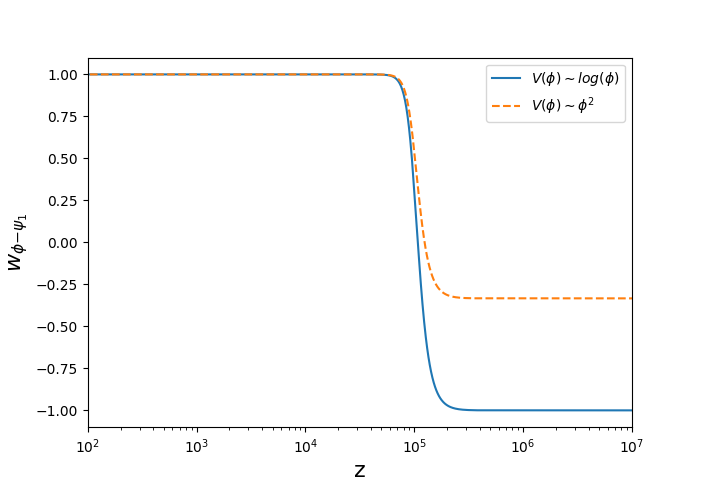}
    \caption{Evolution of equation of state of the scalar fluid for logarithmic and quadratic potentials.}
    \label{fig:my_label}
\end{figure}
  \begin{figure}
\centering
    \includegraphics[scale=0.54]{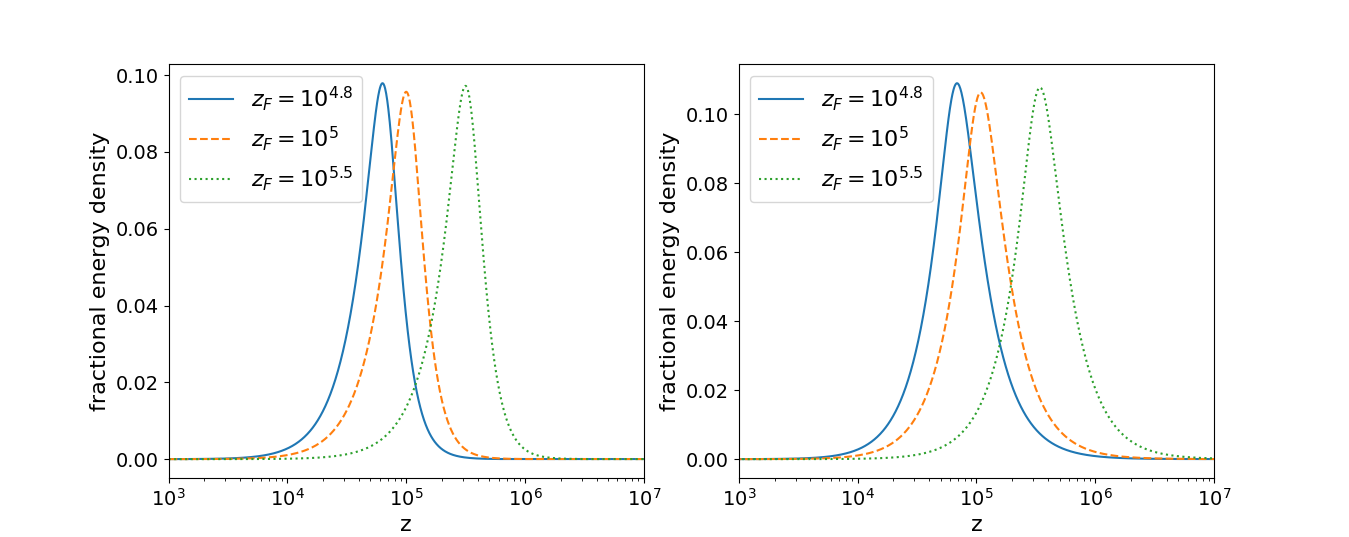}
    
    \caption{Evolution of fractional energy density of early neutrino dark energy for different redshift of nugget formation for logarithmic potential (\textit{left panel}) and for quadratic potential (\textit{right panel}).}
    \label{fractional energy}
\end{figure}

In Fig.~\ref{fractional energy}, we plot the fractional energy density of the fluid against redshift for three different $z_F$ for the cases where w rises from $-1$ to 1 (\textit{left panel}) and from $-1/3$ (\textit{right panel}) to 1 after decoupling around $z_F$. It is shown in \cite{Karwal:2016vyq, Hill:2020osr} that around 5 to 10 percent rise in fractional energy of EDE followed by a quick decay is required to resolve Hubble anomaly. We see a similar behaviour in our numerical solution Fig.~\ref{fractional energy}.

\section{MCMC Analysis}
Here we do a simplest  case MCMC analysis for a neutrino scalar fluid with a quadratic potential. We assume after the phase transition, the kination period dominates and most of the energy of neutrino-EDE dilutes fast as $1/a^6$. Our MCMC analysis indeed shows that this phase transition indeed relaxes the Hubble anomaly.

\pagebreak\subsection{Details Of Analysis }
To perform Monte Carlo Markov Chain (MCMC) analysis, we will be using CMB/BAO  data sets along with SH0ES prior.The Details of our data set are following:
\begin{itemize}
    
    \item Planck 2018 measurements of the low-$\ell$ CMB TT, EE, and  high-$\ell$ TT, TE, EE power spectra, together with the gravitational lensing potential reconstruction \citep{Aghanim:2018eyx}. 
    
     \item The BAO measurements from 6dFGS at $z=0.106$~\citep{Beutler:2011hx}, SDSS DR7 at $z=0.15$~\citep{Ross:2014qpa}, BOSS DR12 at $z=0.38, 0.51$ and $0.61$~\citep{Alam:2016hwk}, and the joint constraints from eBOSS DR14 Ly-$\alpha$ auto-correlation at $z=2.34$~\citep{Agathe:2019vsu} and cross-correlation at $z=2.35$~\citep{Blomqvist:2019rah}.
    
    \item The measurements of the growth function $f\sigma_8(z)$ (FS) from the CMASS and LOWZ galaxy samples of BOSS DR12 at $z = 0.38$, $0.51$, and $0.61$~\citep{Alam:2016hwk}.
    
    \item The Pantheon SNIa catalogue, spanning redshifts $0.01 < z < 2.3$~\citep{Scolnic:2017caz}.
    
    \item The SH0ES result, modelled with a Gaussian likelihood centered on $H_0 = 74.03 \pm 1.42$ km/s/Mpc \citep{Riess:2019cxk}; however, choosing a different value that {\it combines} various direct measurements would not affect the result, given their small differences.

\end{itemize} 
To better understand the extent to which the model can resolve the $H_0$ tension, we perform analysis with  the SH0ES prior as done for most of EDE models in the literature. Our baseline cosmology consists in the following combination of the six $\Lambda$CDM parameters $\{\omega_b,\omega_{\rm cdm},100\times\theta_s,n_s,{\rm ln}(10^{10}A_s),\tau_{\rm reio}\}$, plus two extra parameters of $\nu$EDE , namely $\{\rm Omega_{\rm nuphi}, z_{\rm F}\}$. We dub this model as $\nu$EDE.  We confront the $\Lambda$CDM model with SH0ES prior and  run our MCMCs with the Metropolis-Hasting algorithm as implemented in the MontePython-v3 \citep{Brinckmann:2018cvx} code interfaced with our modified version of CLASS. All reported $\chi^2_{\rm min}$ are obtained with the python package {\sc iMinuit \footnote{\url{https://iminuit.readthedocs.io/}}} \citep{James:1975dr}. We make use of a Choleski decomposition to better handle the large number of nuisance parameters \citep{Lewis:1999bs} and consider chains to be converged with the Gelman-Rubin convergence criterium $R-1<0.05$ \citep{Gelman:1992zz}.

\subsection{Results of MCMC Analysis}
\begin{figure}[hbt!]
\centering
    \includegraphics[scale=0.4]{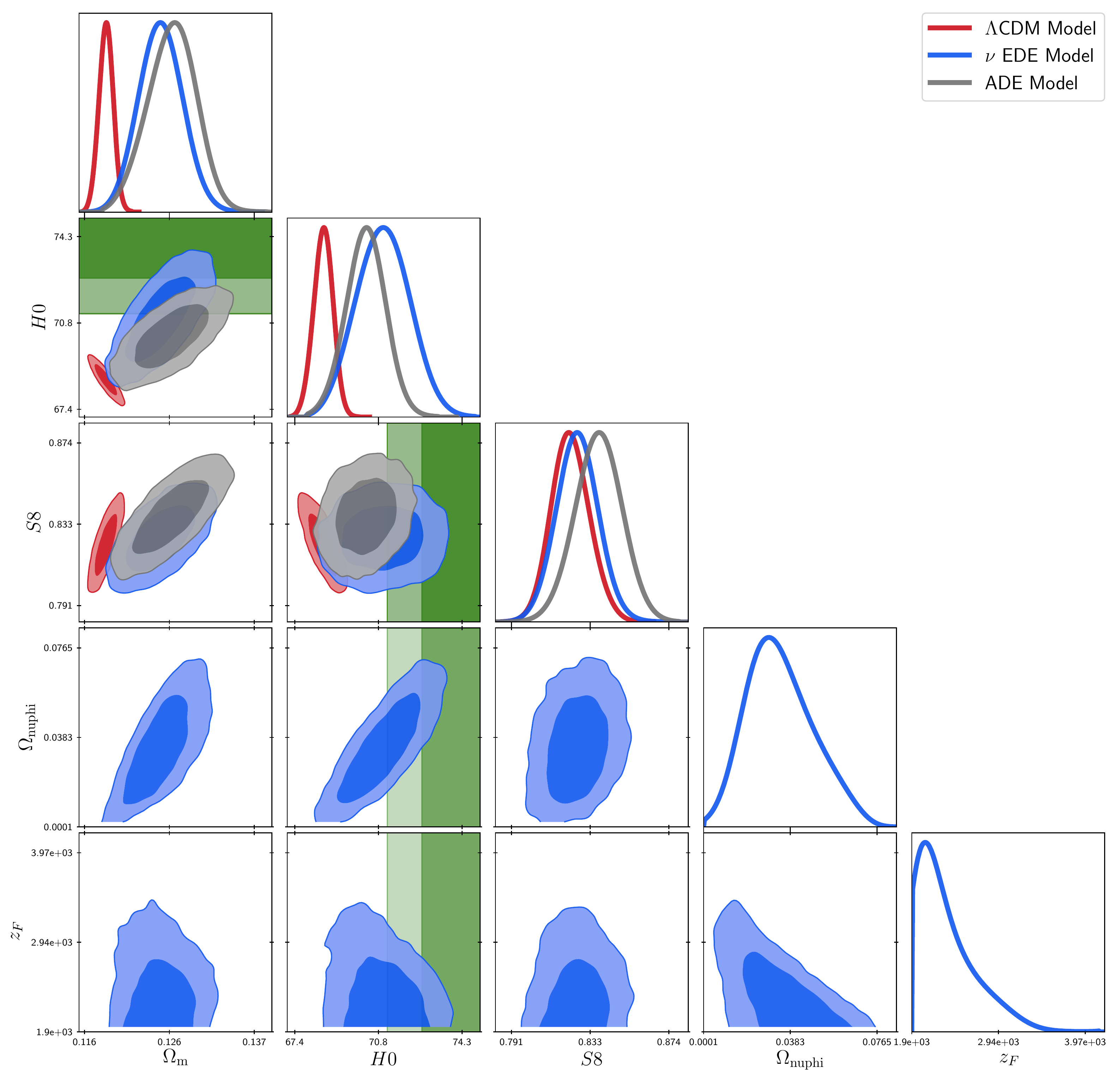}
 \caption{Reconstructed 2D posterior distributions of $(H_0,S_8, \Omega_{\rm m} , \Omega_{nuphi},z_{\rm f})$ is shown for $\Lambda$CDM model vs $\nu$ EDE model vs ADE model with  Planck+BAO/FS+SN1a+SH0ES data .We also have added 68\%(Dark green) and 95\%(light green) bands corresponding to a gaussian H0 prior/likelihood Values from \citep{Riess:2019cxk}. \label{fig:MCMC}}
    
\end{figure}

The parameter constraints obtained from $\nu$EDE and $\Lambda$CDM using Planck+Ext+SH0ES data are shown in table \ref{chi2_EDE}. Where 'Ext' represents combined '$\rm {BAO/FS+SN1a}$'. We also have reported $\chi_{\rm min}^2$ for each model in table \ref{chi2_EDE}.
The 2d posterior distributions for $\nu$EDE along with $\Lambda$CDM and acoutic dark energy (ADE) model are shown in figure \ref{fig:MCMC} .\\
From the results, we have evidence for non-zero $\rm Omega_{\rm nuphi}$at a epoch of transition $z_{\rm F}$. To be precise we have got bestfit of $\rm Omega_{\rm nuphi}$=0.05631 and redshift $z_{\rm f}$=1903.7 resulting a significantly higher value of Hubble parameter $H_0=71.017(71.69)_{-1.144}^{1.123}$.
With this value of Hubble parameter, we are in agreement of nearly 1.3$\sigma$ with SH0ES measurement which is $H_0 = 74.03 \pm 1.42$ km/s/Mpc \citep{Riess:2019cxk}. For better visualization please refer to  the figure \ref{fig:MCMC}, where we have shown 1$\sigma$(dark green) and 2$\sigma$(light green) bands of SH0ES measurements of Hubble parameter.  In terms of the goodness of fit, $\chi_{min}^2$ numbers that we have shown in table \ref{chi2_EDE}. We  got an improvement of  $\Delta \chi_{tot}^2$=-10.51 which shows the  Planck+Ext+SH0ES data prefers $\nu$EDE cosmology over  $\Lambda$CDM cosmology.

If we compare Our results of $\nu$ EDE cosmology with Accoustic Dark Energy(ADE) cosmology \citep{Lin:2019qug}, our $\nu$ EDE cosmology is doing marginally better in terms of agreement with SH0ES measurement and  measurements  of  the  amplitude  of  local  structure. As we can see we get a smaller $S_8$ than ADE models. Detailed Analysis will be done  separately in near future using a different  cosmological tool \citep{DAmico:2020kxu} and it is beyond the scope of this paper.

\begin{table*}[hbt!]
\centering
\scalebox{0.9}{
  \begin{tabular}{|l|c|c|}
     \hline
     Model & \multicolumn{1}{c|}{$\Lambda$CDM} & \multicolumn{1}{c|}{Early mass varying neutrino DE} \\
     \hline Parameter & Planck+Ext+SH0ES&Planck+Ext+SH0ES \\ \hline \hline
      $100~\theta_s$ & $1.0421(1.04204)_{-0.000284}^{+0.000288}$ &$1.0412(1.04107)_{-0.000414}^{+0.000417}$ \\
    $100~\omega_b$  & $2.252(2.247)_{-0.01352}^{0.01309}$&$2.2759(2.263)_{-0.02054}^{  +0.01703}$\\
    $\omega_{\rm cdm}$&$0.1184(0.1191)_{-0.000895}^{+0.000878}$&$0.12535e(0.1254)_{-0.002811}^{+0.002563}$ \\
    ${\rm ln}(10^{10}A_s)$ &$3.053(3.0443)_{-0.01525}^{+0.01422}$& $3.0586(3.0574)_{-0.01529}^{+0.01423}$   \\
    $n_s$&$0.9695(0.9676)_{-0.003814}^{0.003679}$&$0.9740(0.9732)_{-0.00462}^{+0.00434}$  \\
    $\tau_{\rm reio}$ &$0.05984(0.05438)_{-0.00791}^{+0.00707}$& $0.03432(0.05631)_{-0.01719}^{+0.01337}$ \\
    $z_{\rm f}$  & $-$ &2$338.1(1903.74)_{-438.09}^{+107.64}$\\
     $\Omega_{\rm nuphi}$  & $-$&$0.029275(0.05308)_{-0.01715}^{+0.01095}$\\
    \hline
     $H_0$ [km/s/Mpc] & $68.16(67.83)_{-0.409}^{+0.393}$& $71.017(71.69)_{-1.123}^{+1.144}$  \\
    \hline
    $ \chi^2_{\rm tot}$  & $3830.41$ & $3819.90$\\
    \hline
    $\Delta \chi^2_{\rm min}$ ($\Lambda$CDM) &0 &-10.51\\
    \hline
  \end{tabular}
  }
  \caption{The mean (best-fit) $\pm1\sigma$ error of the cosmological parameters reconstructed from the lensing-marginalized Planck+BAO+SN1a+SH0ES data. We also report the $\Delta \chi^2_{\rm min}$ for $\nu$EDE  model.}
  \label{chi2_EDE}
\end{table*}

  But unlike other EDE model, our DE scale is not fine tuned rather EDE  emerges naturally during MRE because of similarity of  light sterile neutrino mass and  Temperature of Universe at around matter radiation equality. So  our model is a  non-fine tuned EDE model which also statistically preferred over LCDM  and resembles EDE or acoustic DE models. As,  building a viable EDE model without fine-tuning of dark energy scale is very difficult, we feel, our neutrino mass driven $\nu$EDE is a step forward for a  successful non- fine tuned model of early dark energy and it's worth exploring in details as future research.
  It is true that like other EDE or Acoustic DE model, our models  is also challenged by LSS data as shown in \cite{Hill:2020osr} . But our $S_8$  value is slightly smaller than standard ADE models. On top  there are two points we would like to mention.

\begin{itemize} 
  
 \item {Recently, after \cite{Hill:2020osr}, there has been another work \cite{smith2020early}  where they have shown that EDE models are constrained but  not ruled out by LSS data-specially galaxy power spectrum measuremnt aspect of LSS is fine with EDE models while recent weak lensing KiDS/Viking \cite{Joudaki_2020,Hildebrandt_2020} put strong constraints on it.  So, EDE models are still an active area of research  and we feel near future LSS data will  determine its fate. But as seen from initial MCMC run, our model somehow does not make $S_8$ worse like EDE or ADE models as seen in figure \ref{fig:MCMC}.}
 
\item  {On top, our model has an extra feature which we  guess, will   help us when confronted with LSS data in future.
 The nuggets can evaporate in scalar radiation as mentioned in \cite{Afshordi:2005ym}. That will make the gravitational potential decay and suppress the matter power spectra and might alleviate LSS difficulty completely. It was shown in \cite{Pandey:2019plg} that even 1 or 2 \% of DM decays into radiation after decoupling, it can alleviate $S_8$ tension significantly. 
 How much radiation the nuggets produce and depending on  whether they are thermal or non-thermal bath, it will modify our matter power spectra. This needs detailed analysis and we hope to report this aspect in near future}.\\
 \end{itemize}

\section{Conclusion} \label{sec:conclusion}
Data from MiniBooNE experiment might indicate the existence of  light sterile neutrino states of mass (sub-\rm{eV} to 10 \rm{eV}). Though this lighter sterile neutrino has relevant properties, in general, light \rm{eV}-mass sterile fermions are not viable cold dark matter (CDM) candidate due to its excessive free-streaming. 
 Here in this paper 
 
 \begin{itemize}
     \item 
 We realize  a scenario of nugget formation with $\rm{eV}$ sterile neutrino  in radiation dominated era close to MRE. We analytically show that the condition of instability is generic even with quadratic potential in RDE when sterile neutrino turns non-relativistic followed by nugget formation. Our scenario provides a complete new way of producing a cold dark matter nuggets from light \rm{eV} sterile neutrino. This dark matter can contribute to a considerable fraction of DM density today depending on model paramters.
 \end{itemize}
 
 \begin{itemize}
     \item 
 Using Thomas Fermi approximation and solving coupled Klein Gordon equation numerically, we find the static configuration of the scalar field which allows us to calculate nugget mass, radius of nugget and dark matter density. The detailed nugget profile  was never solved  before in  neutrino DE model. The mass of the nugget in our two examples are 
 $\sim 10^{6}$ eV and $\sim 10^{29}$ eV and the corresponding redshift of nugget formation are $z_F \sim 0.86 \times 10^5$ and $6 \times 10^7 $ respectively. For these formation redshifts , we show that the entire dark matter density of the university can arise from this heavy neutrino nuggets!
 \end{itemize}

 \begin{itemize}
     \item 
 We use quadratic scalar potential for the scalar field as it keeps the radiative correction to the potential  under control as illustrated in super-symmetric theories of neutrino dark energy \citep{Fardon:2005wc}. Also we use most general Yukawa type  coupling in dark sector as proposed in original neutrino dark energy model.
 \end{itemize}
 
  \begin{itemize}
     \item 
 
  This natural phase transition prior to MRE in  Early neutrino dark energy has strong implications for recent Hubble tension. This model of EDE does not require a fine-tuned dark energy scale yet it can comprise around 10 percent of total energy budget around MRE ( as needed to solve Hubble anomaly \citep{Poulin:2016nat,Poulin:2018cxd}). This early dark energy  decays faster than radiation as the nugget forms and scalar rolls of freely along the quadratic potential dominated by a kinetic energy phase. In our case, the DE energy injection and its decay ( due to phase transition) is entirely controlled by  $\rm{eV}$ sterile neutrino mass and in principle our scenario  can realise a viable early DE particle physics model which can solve Hubble tension. EDE models are still an active area of research  and we feel near future LSS data will  determine its fate. But as seen from initial MCMC run, our model is consistent with higher $H_0$ and does not make $S_8$ worse like EDE or ADE models as seen in figure \ref{fig:MCMC}. A detailed LSS data analysis is required to precisely tell how well our model performs and what kind of potential of scalar field is preferred when confronted with large scale structure data from galaxy survey or recent  weak lensing measurements. This will be reported in near future. \\
  \end{itemize}       

\subsection*{Acknowledgements} 
We thank Vivian Poulin and Ethan Nadler for giving us comments on the draft. We also thank Marc Kamionkowski for suggesting us to adopt generalized dark matter formalism for neutrino-scalar fluid and look for instability in terms of sound speed. We are grateful to  James Unwin for reading the manuscript and giving us valuable suggestions. We thank Neal Weiner and Kris Sigurdson for helpful discussions during the initial phase of the work.  SD thanks  IUSSTF-JC-009-2016 award from the Indo-US Science \& Technology Forum which supported the project.



\bibliography{apj_draft}{}
\bibliographystyle{aasjournal}

\end{document}